\documentclass[oldversion]{aa}

\usepackage{epsfig}
\usepackage{graphics}
\usepackage{float}
\usepackage{amsmath}
\usepackage{multirow}
\usepackage{longtable}
\usepackage{rotate}
\usepackage{array}
\usepackage{subfigure}
\def\kms{\ifmmode{\rm km\,s^{-1}}\else\hbox{$\rm km\,s^{-1}$}\fi}

\setlongtables

\begin{document}

\title{Mass fluxes for O stars}

\author{L.B.Lucy}
\offprints{L.B.Lucy}

\institute{Astrophysics Group, Blackett Laboratory, Imperial College 
London, Prince Consort Road, London SW7 2AZ, UK}
\date{Received ; Accepted }

\abstract{The theory of moving reversing layers for hot stars
is updated to include
an extensive line list, a radiative boundary condition from
static model atmospheres, line
transfer by scattering, and continuation to supersonic velocities. A Monte 
Carlo technique determines the theory's eigenvalue $J$,
the mass flux, and the derived $J$'s are in good agreement with the wind models
of Pauldrach et al. (2001).
The solutions' sensitivity to the photospheric microturbulent velocity
$v_{t}$ reveals that this parameter has a throttling
effect on $J$: turbulent line-broadening in the quasi-static layers reduces
the radiation force available to accelerate matter through the sonic point.
If photospheric turbulence 
approaches sonic velocities, this mechanism reduces mass loss rates by factors
$\ga 3$, which would partly account for the reduced rates found observationally
for clumpy winds. 
}

\authorrunning{Lucy}
\titlerunning{Mass fluxes}
\maketitle

\section{Introduction}

The line-driven winds of hot stars are unstable (Lucy \& Solomon 1970,
[LS70]; Owocki et al. 1988) and the resulting density
inhomogeneities strengthen
emission lines such as H${\alpha}$ and He$\:${\sc ii} $\lambda$4686,
thus positively biasing empirical estimates of $\Phi$, the mass-loss rate
(Lucy 1975; Abbott et al. 1981). This bias could be corrected
if the
statistical properties of the inhomogeneities could be predicted, but this
is still not feasible.
In consequence, clumpiness is now an essential extra
parameter in diagnostic analyses of the supersonic outflows from early-type
stars (Schmutz 1997; Hillier \& Miller 1999).  
That such investigations confirm clumpiness is
no great surprise in view of work in the 1980's concerning
X-ray emission, the black troughs of saturated P Cygni lines, and the 
variability of line profiles. But the magnitude of the
concomitant reductions of the inferred $\Phi$'s is a major surprise,
as is the implication that theory overpredicts $\Phi$'s by substantial factors.

The $\Phi$'s of Galactic O stars are investigated in the
recent papers of Bouret et al. (2005) and Fullerton et al. (2006), both of
which also review earlier literature. Bouret et al. analyze the spectra of two
O4 stars and derive $\Phi$'s that are factors of 3 and 7 below predicted
values. On this basis, they conclude that our understanding
of mass loss by O stars is in need of fundamental revision.

Fullerton et al. study a much larger sample of 40 stars  
but, instead of analysing entire spectra, they focus on the P$\:${\sc v} 
resonance doublet which, being seldom saturated, more readily
yields reliable $\Phi$'s than do lines from abundant species.
Strong clumping, reducing previous empirical $\Phi$'s by factors of 10 or more,
is required to match the rates derived from P$\:${\sc v}. These authors
therefore conclude that the ``standard model'' for hot-star winds needs to be
amended.

The theoretical $\Phi$'s tested by Bouret et al. (2005) are those
derived by Vink et al. (2000) with a refined version 
of the Monte Carlo (MC) technique of Abbott \& Lucy
(1985,[AL85]).
As such, 
these $\Phi$'s are not eigenvalues of the differential equations governing 
outflows but are
semi-empirical estimates based on an assumed wind structure. 
Accordingly, if these $\Phi$'s are 
in conflict with observation, the fault may lie in the adopted stratification
rather than in the underlying theory.  

At present, the closest to a complete theory of radiatively-driven winds is
embodied in the
code WM-{\em basic} of Pauldrach et al. (2001). Thus, for a critical
test of wind theory, the above
reduced empirical $\Phi$'s should be compared to the eigenvalues $\Phi$ of
WM-{\em basic} models 
tailored to individual stars. Pending this, some progress is possible with the 
models in Table 5 of Pauldrach et al. (2001). Interpolation 
using the Bouret et al. (2005) parameters for 
HD 190429A gives a WM-{\em basic}
$\Phi \simeq 6.3 \times 10^{-6} \sun \; yr^{-1}$, a factor 3.5 larger than
their clumped-wind model. Thus
claims that line-driven wind theory is in need of major revision appear to
be well founded.    

Recent papers on clumpiness confirm what was already evident
in the mid 1980's, namely that the ``standard model'' of line-driven winds
must eventually be replaced by 3-D time-dependent gas-dynamical models with
simultaneous NLTE radiative transfer. Such models are necessary if we
wish to
predict from first principles spectral features formed at supersonic
velocities.
But if clumping is negligible in the neighbourhood of the sonic point,
the standard model's predictions of $\Phi$ should be fairly accurate since
the differential effect of clumpiness at supersonic velocities on the radiation
field at the sonic point is surely slight. Thus the evidence that the
predicted
$\Phi$'s are far from accurate perhaps indicates that clumpiness is already
significant at the sonic point, a conclusion with some observational
support (Bouret et al. 2005).

Notwithstanding this latter possibility, this paper is motivated by the belief
that the eigenvalues $\Phi$ of
stationary-wind solutions should indeed be fairly accurate.
Accordingly, the moving reversing layer model of LS70 is revived in order to
compute mass fluxes for O star atmospheres and to investigate their
sensitivity to assumptions and parameters.

\section{An outflowing reversing layer}

In this section, the LS70 model of stationary flow 
accelerating to sonic velocity is updated.
The major changes are a vast increase in the line list, a realistic lower
boundary condition on the radiation field, an improved treatment of line
formation, and continuation to supersonic velocities. The solution procedure
is also changed. Because line formation by
pure absorption is replaced by scattering, solution with initial-value
integrations is no longer possible. Accordingly, a MC technique akin to
that of AL85 is developed.

Despite these improvements, the resulting code falls far short of the physical
detail and numerical precision achieved by codes such as WM-{\em basic},
TLUSTY (Lanz \& Hubeny 2003) ,
CMFGEN  (Hillier \& Miller 1998), and FASTWIND (Puls et al. 2005).
But high accuracy is surely not needed in identifying an effect capable of
reducing $\Phi$ by factors $\ga 3$.

\subsection{The model} 
  
As in LS70, a Schuster-Schwarzschild model is adopted.
Thus the lower boundary is a continuum-emitting surface above which is
an outflowing plane-parallel isothermal layer where only line- and electron
scatterings
occur. In addition to chemical composition, the basic parameters are
effective temperature $T_{eff}$,
surface gravity $g$, microturbulent velocity $v_{t}$, and mass flux 
$J = \Phi/4 \pi R^{2}$.

The composition is solar with $N_{He}/N_{H}=0.1$ (Grevess \& Sauval 1998),
with the included ions as in Table 1 of Lanz \& Hubeny (2003).
The temperature of
the isothermal flow is set $ = 0.75 T_{eff}$, and 
$H$ and $He$ are assumed to be fully ionized, so that
thermodynamic coefficients are height-independent.

\subsection{Gas dynamics} 

The structure of the reversing layer is determined by the equations
governing stationary isothermal flow. With sphericity neglected, the 
continuity equation has the integral 
\begin{equation}
    \rho v = J
\end{equation}
and the equation of motion can then be written as
\begin{equation}
    (v^{2}- a^{2}) \: \frac{1}{v} \frac{d v}{d x} = -g_{eff}
\end{equation}
Here $a = \sqrt{ P/\rho}$ is the isothermal speed of sound and the effective
gravity
\begin{equation}
    g_{eff}= g - g_{e} - g_{\ell}  
\end{equation}
where $g_{e}$ and $g_{\ell}$ are the radiative accelerations due to electron-
and line scatterings, respectively.

From Eq.(2), we see that the sonic point $v = a$ is a critical point at which
the velocity gradient is finite only if $g_{eff} = 0$. This condition was used 
in LS70 to determine the eigenvalue $J$.

Subsequent to LS70, Castor, Abbott \& Klein (1975, [CAK]) pioneered the
development
of a complete theory of stationary line-driven winds by exploiting the
computational economy of the Sobolev approximation. However, 
the critical point is then no longer at the sonic point but is displaced 
downstream into the supersonic flow. Here, as in LS70,
the Sobolev approximation is not employed. In consequence, $g_{\ell}$ is
a functional of the solution, and so the critical point remains at the
sonic point (Lucy 1975; Poe et al. 1990).

\subsection{Stratification}

In AL85 and Vink et al. (2000), the MC transfer is
performed in an observationally-motivated kinematic model, with 
$\Phi$ determined iteratively by imposing a global
dynamical constraint. Here, with the aim of achieving greater local
consistency, a family of dynamical models is generated
corresponding to an assumed functional form for $g_{\ell}(v)$.  
The mass flux $J$ is then determined iteratively by bringing
$g_{\ell}(v)$ and its MC estimator
$\tilde{g}_{\ell}(v)$ into approximate agreement. 

The adopted $g_{\ell}(v)$ is the two-parameter formula
\begin{equation}
   g_{\ell} = g_{*} max \left[\delta, \left(\frac{v}{a}\right)^{s} \right]
      \;\;\; with \;\;\; g_{*} = g - g_{e}
\end{equation}
With $0 < \delta < 1$ and $s > 0$, this formula is such that $g_{eff} = 0$ at
$v = a$ and is negative at supersonic velocities. 
The discontinuous derivative of this function at the join of its two
segments allows the sharp increase in $g_{\ell}$
as the outflow accelerates up to and through the sonic point to be captured. 

The solution of Eq. (2) that avoids a singularity at the sonic point has
\begin{equation}
  \left( v \frac{dv}{dx} \right)_{a} = 
        \frac{1}{2}\: \left( \frac{d \ell n g}{d \ell n v} \right)_{a} g_{*}
 \;  =  \frac{1}{2}\: s \: g_{*}
\end{equation}
Accordingly, the velocity law $v(x;\delta,s)$ is derived from Eq. (2) by
inward and outward integrations from $v = a$, each with initial derivative
given by Eq. (5).

Together with Eq. (1) and the isothermal assumption, $v(x;\delta,s)$ determines
the stratification throughout the computational domain 
$(x_{0},x_{1})$. 
The lower boundary $x_{0}$ is at electron-scattering optical
depth $\tau_{e} = 2/3$ below the sonic point. The upper boundary $x_{1}$
is at $v \sim 4a$. 

Note that, as in solar-wind theory, the solution has an X-type 
topology at the critical point. 
This is assumed without proof because of the difficulty of analysing the
solution topology when $g_{\ell}$ includes contributions from
scattered radiation (Poe et al. 1990).

\subsection{Radiation field}

The frequency distribution of the continuum intensity $I_{\nu}^{+}$ emitted
at $x_{0}$ is that of the emergent {\em continuum} flux of
the TLUSTY (Lanz \& Hubeny 2003) solar-abundance model of the specified
$T_{eff}$ and $g$; the ionization edges due to H and He are therefore
accurately represented. The angular distribution 
is assumed uniform. 

At $x_{1}$, the integrated flux of the
outwardly-directed radiation $I_{\nu}^{+}(\mu)$ is constrained to 
$= \sigma T_{eff}^{4}$. Correspondingly, $I_{\nu}^{-} = 0$ at $x_{1}$, so
that back-scattering from the highly supersonic flow is neglected. 
Nevertheless, $x_{1}$  
is far enough downstream that, for $v \simeq a$, each line's 
self-interactions with back-scatterings from $v>a$ are treated.

As it propagates through the reversing layer, the continuum is
modified by electron scatterings and line transitions. 
Electron scatterings are assumed to be isotropic but frequency shifts
in the co-moving frame (cmf) due to the electrons' thermal motions
are taken into account.
Following Sect. IIf in AL85, line transitions are treated as resonance
scatterings. These are assumed to be isotropic with a
re-emitted frequency in the cmf in accordance with complete
redistribution. The line profile $\phi_{\nu}$ comprises a Doppler core due to
thermal and microturbulent motions and wings due to
radiation damping.

\subsection{Ionization and excitation}
   
The treatment of ionization and excitation is designed to give moderately
accurate populations for the lower levels of transitions that are major
contributors
to $g_{\ell}$. Thus the emphasis is on the ground states and low-lying
metastable levels of abundant ions. Conveniently, NLTE departure
coefficients for O-star atmospheres can be downloaded from
the TLUSTY website, and data from a model with $T_{eff} = 50000K$,
$log g = 4$ are plotted in Fig. 1 for the important ion Fe$\:${\sc v}.
This diagram
shows the expected convergence to rigorous LTE in deep layers and the
growing importance of NLTE in the surface layers. However,
because of collisional coupling, the low-lying superlevels of  
Fe$\:${\sc v} remain approximately in LTE relative to the ground state.
In contrast, for the ionization ratio
Fe$\:${\sc v}/Fe$\:${\sc iv}, significant departures from LTE occur for 
$n_{e} \la 13$ dex. Nevertheless, changes in $n_{e}$ remain 
the dominant cause of ionization gradients in the surface layers.

In view of the above, excited levels are assumed to be in LTE with their
ground states, whose departure coefficients are fixed at the
TLUSTY values in a representative layer. Thus, the
ionization equation is 
\begin{equation}
   \frac{n_{J+1}n_{e}}{n_{J}} = \left(\frac{b_{1,J+1}}{b_{1,J}}\right)
   _{\dagger} \left(\frac{n_{J+1}n_{e}}{n_{J}}\right)^{*}_{\dagger}
\end{equation}
where $*$ denotes LTE and $\dagger$ indicates values for the layer
with $T_{e} \simeq 0.75 T_{eff}$.

\begin{figure}
\vspace{8.2cm}
\includegraphics{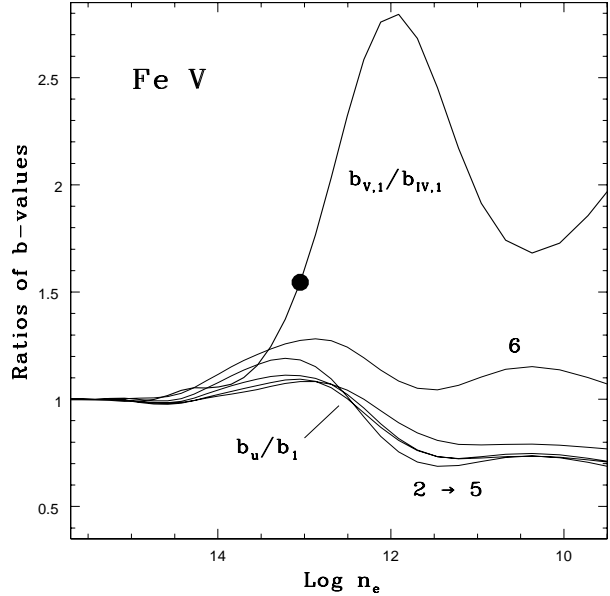}
\caption{Ratios of departure coefficients relative to the ground state for
selected levels of Fe as functions of $n_{e}$ for TLUSTY model
G50000g400v10.12. Ratios are plotted for superlevels of Fe$\:${\sc v} 
with excitation energies $< 10$ eV. Also plotted is the ratio of b-values for
the ground states of Fe$\:${\sc v} and Fe$\:${\sc iv}. The filled circle is
the point where $T_{e} \simeq 0.75 T_{eff}$.}
\end{figure}

\subsection{Line list}

The basic line data is from Kurucz \& Bell (1995).
A typical model treats the line
formation of $\sim 10^{5}$ lines in the wavelength interval
228 - 4000\AA\ .

\section{Monte Carlo code}

In this section, a MC code is described for calculating 
$g_{\ell}$ as a function of height in the outflowing reversing layer
defined in Sect. 2. The description is brief, focussing on aspects
not covered in AL85.

\subsection{Discretization}

The continuous model derived in Sect.2 is approximated by $\sim 100$
slabs, within each of which $\rho$ and $v$ are constant. The velocity increment
between neighbouring slabs is everywhere $\ll v_{t}$.

\subsection{Scattering histories}

A total of $\cal N$ photon packets of energy $\epsilon_{0}$ are launched into
the computational
domain at $x_{0}$ in time $\Delta t$. Their initial rest frame frequencies
$\nu$ sample the continuum
flux distribution of the TLUSTY model and their initial direction
cosines are given by $\mu = \sqrt{z}$, where here and below $z$ denotes
a random uniform variate obtained with ran2 (Press et al. 1992). 

A packet's straight-line trajectory in a slab terminates after
travelling distance $s = min(s_{b}, s_{e}, s_{\ell})$, where $s_{b}$ is
the distance to the slab's edge, $s_{e} = -\ell n (z)/\sigma_{e}$, and  
$s_{\ell} = -\ell n (z)/\ell_{\nu'}$. Here $\sigma_{e}$ and $\ell_{\nu'}$
are, respectively, the electron- and line scattering coefficients per unit
volume, the latter evaluated at $\nu'$, the packet's cmf frequency. Note
that these coefficients are constant within each slab.   

If the packet undergoes an electron scattering, its new 
$\mu = 1-2z$ and its new $\nu'$ is obtained by sampling
the angle-averaged redistribution function (Mihalas 1978, p.432). 
For a line scattering, the new 
$\mu = 1-2z$ and the new $\nu'$ is obtained from
the assumption of complete redistribution.

The scattering history of a packet is complete when it exits the
computational domain at $x_{0}$ or $x_{1}$. The number ${\cal N}^{+}$
escaping at $x_{1}$ constitute the 
emergent radiation and so must have an integrated flux corresponding to
$T_{eff}$.
Thus, at the completion of the MC experiment, the packets'
energy $\epsilon_{0}$ is given by   
\begin{equation}
       {\cal N}^{+} \: \frac{\epsilon_{0}}{\Delta t} = \sigma \: T_{eff}^{4}
\end{equation}

\subsection{Estimator for $g_{\ell}$}

With $\epsilon_{0}/ \Delta t$ from Eq. (7), the MC 
radiation
field, comprising the trajectories of $\cal N$
packets, can be transformed to physical units and then used to
calculate $g_{\ell}$.

If the intensity $I_{\nu}(x,\mu)$ were known, $g_{\ell}$ could be
computed from the formula
\begin{equation}
  \rho g_{\ell} = \frac{2 \pi}{c} \int_{-1}^{1}\int_{0}^{\infty} \ell_{\nu'}
                        I_{\nu} \mu \: d \mu \: d \nu
\end{equation}
valid for  $v \ll c$. This can be applied to the energy-packet model using the
intensity estimator
\begin{equation}
 \tilde{I}_{\nu} \: d \nu \: d \omega = \frac{\epsilon_{0}}{\Delta t}
   \frac{1}{V} \sum_{ d \nu d \omega} \frac{\epsilon_{\nu}}{\epsilon_{0}} \: s
\end{equation}
where the summation is over all pathlength segments $s$ in the reference volume
$V$ of packets in
$(\nu, \nu + d \nu)$ propagating in solid angle element $d \omega$. This
formula
is derived in Lucy (2005) and repeated in order to correct a typographical
error. Eq. (9) relates the energy-packet and the specific-intensity
representations of radiation. 

In the present application, $v \ll c$ and so setting
$\epsilon_{\nu} = \epsilon_{0}$ is an excellent approximation. Also the
geometry is plane-parallel, so $V$ is a column of unit 
cross-section and length $\Delta x$, the slab's width. Substitution into
Eq. (8) then gives the estimator
\begin{equation}
  \rho \tilde{g}_{\ell} = \frac{1}{c} \: \frac{\epsilon_{0}}{\Delta t} \:
        \frac{1}{\Delta x} \sum \ell_{\nu'} \: \mu \: s
\end{equation}
where the summation is over all trajectories in the slab.

This optimum estimator for $g_{\ell}$ exploits all the information 
in the packets' trajectories. This would not be the case
if $g_{\ell}$ were derived from the momentum transfers occurring when  
packets undergo line scatterings. Even if no line scatterings occur in a
slab, Eq.(10) returns a non-zero 
$\tilde{g}_{\ell}$, provided only that at least one packet propagates with
a $\nu'$ at which $\ell_{\nu'} \neq 0$.

Because this calculation of ${g}_{\ell}$ does not rely on the Sobolev
approximation, the effects of scattered radiation discussed by 
Owocki \& Puls (1999) are included.
Moreover, line-profile overlapping and multiline transfer are accurately
treated. These effects are excluded when Sobolev theory is used in the
narrow-line limit with the
assumption that each line interacts with unattenuated continuum.

\subsection{Computational economy}

With the procedure as described,
a prohibitive amount of computer time would be required for an
accurate $\tilde{g}_{\ell}$. Accordingly, two
simplifications are introduced.

A MC treatment of re-emission with complete redistribution should
first randomly select which of the overlapping lines contributing to  
$\ell_{\nu'}$ is the actual absorber and then sample its 
profile $\phi_{\nu'}$ to select the $\nu'$ of the emitted packet. This has
been simplified by always taking the absorber to be the largest contributor to
$\ell_{\nu'}$, which is identified when $\ell_{\nu'}$ is tabulated prior to
the MC calculation. Little loss of accuracy is expected since
absorption usually occurs in a line's Doppler core and this typically
strongly dominates the damping wings of neighbouring transitions. 

The second simplification avoids trapping in the Doppler cores
of strong lines.
A photon undergoing numerous scatterings in a Doppler core is described by
Rybicki \& Hummer (1969)
as being imprisoned at a point and from which it 
escapes only when the low probability event of emission in the damping wings
occurs. When a MC packet is similarly imprisoned, its escape via
the damping wings is expedited. 
The actual procedure is as follows: with
standard notation, the
Voigt profile is approximated (Mihalas 1978, p.281) by the gaussian core when  
$|v| < v^{*}$ and by the damping wings when $|v| > v^{*}$, where $v^{*}$ is
such that the two terms are equal. If a slab is optically thick in a Doppler
core, the next re-emission is forced to occur in the wings. Since 
$\phi_{\nu} \propto 1/v^{2}$ in the wings, a random $v = \pm v^{*}/z$.

\section{Examples}  
 
In this section, moving reversing layers are derived for the three TLUSTY
atmospheres with parameters matching those of 
models in Table 5 of Pauldrach et al. (2001).
Mass fluxes can then be compared without interpolation, thereby testing
the approximations used to construct the simple
code described in Sects. 2 and 3.
 
\subsection{Parameter fitting}  

With $T_{eff},g$ and composition fixed, the remaining parameters are
$J, \delta, s$ and $v_{t}$. 
In the WM-{\em basic} models, $v_{t}$
increases from 10 km s$^{-1}$ at the photosphere to $0.1v_{\infty}$
in the terminal flow. Here, we ignore 
this gradient and set
$v_{t} = 10$ km s$^{-1}$.

With a trial parameter vector ($J, \delta, s$), a MC calculation is carried
out as described in Sect. 3.
Eq.(10) then gives $\tilde{g}_{\ell}$ throughout the reversing layer and thus
allows us to see how closely the assumed $g_{\ell}(v)$ given by Eq. (4)
is recovered. The goodness-of-fit
is quantified by calculating
\begin{equation}
  \eta = \frac{1}{I} \sum \left|\frac{ \tilde{g}_{\ell}-g_{\ell}}{g_{\ell}}
                                                                 \right|
\end{equation}
where the summation is over the $I$ slabs with $0.25 < v/a < 1.75$. 
Velocities
outside this interval are excluded to avoid the edge effects  
due to imperfect boundary conditions.

With many repetitions of this procedure, the vector $(J, \delta, s)$
minimizing $\eta$ can be found.
In the initial search, models are calculated with
${\cal N} = 10^{7}$, but this is increased to $4 \times 10^{7}$ in the close
neighbourhood of the minimum and then further increased to    
${\cal N} \ga 10^{8}$ for the illustrations herein.

Since the interval $(0.25,1.75)$ includes the sonic point, the regularity
condition at $v=a$ is satisfied exactly if $\eta = 0$. But because of MC
sampling errors
and the limited flexibility of the formula for $g_{\ell}$, this is not
achievable. Accordingly,  
minimizing $\eta$ replaces the condition $g_{eff} = 0$ at $v = a$ as the
criterion determining the eigenvalue $J$.

\subsection{Model D-50}  

Model D-50 of Pauldrach et al. (2001) has 
$T_{eff} = 50000$K, $log \: g = 4$, $R/R_{\sun} = 12$, and 
$\Phi = 5.6 \times 10^{-6} M_{\sun}/yr$, corresponding to a photospheric
mass flux $J(gm \: s^{-1})  = -4.40$ dex .
The matching TLUSTY model is G50000g400v10.12, details of which were
used for Fig. 1. From the emergent spectrum of this model, continuum fluxes
were estimated at intervals of 0.1 in $log \nu$ and at both
sides
of ionization edges. Linear logarithmic interpolation then defines the
$\nu$-distribution of $I_{\nu}^{+}$ at $x_{0}$.

The best-fit moving reversing layer has parameter vector
$(log J, \delta, s) = (-4.41,0.61,2.03)$
which gives $\eta = 0.048$, and this
fit worsens with displacement
vectors $(\pm 0.05, \pm 0.03, \pm 0.05)$. Thus the mass flux of this model
agrees with the WM-{\em basic} model to within 0.05 dex.

The success in reproducing the assumed $g_{\ell}$ is shown in Fig. 2. The
agreement is seen to be excellent from $v \simeq 10$ to
$\simeq 70$ km s$^{-1}$,
an interval containing the sonic point at $22.6$ km s$^{-1}$. In
particular, the 
discontinuity in the derivative of  $g_{\ell}(v)$ at
$v = 17.5$ km s$^{-1}$ closely matches a sharp change in the slope of
$\tilde{g}_{\ell}(v)$. On the other hand, the fit is less successful 
for $v \la 10$ km s$^{-1}$, where Eq. (4) reduces to
$g_{\ell} =  g_{*} \delta$.

\begin{figure}
\vspace{8.2cm}
\includegraphics{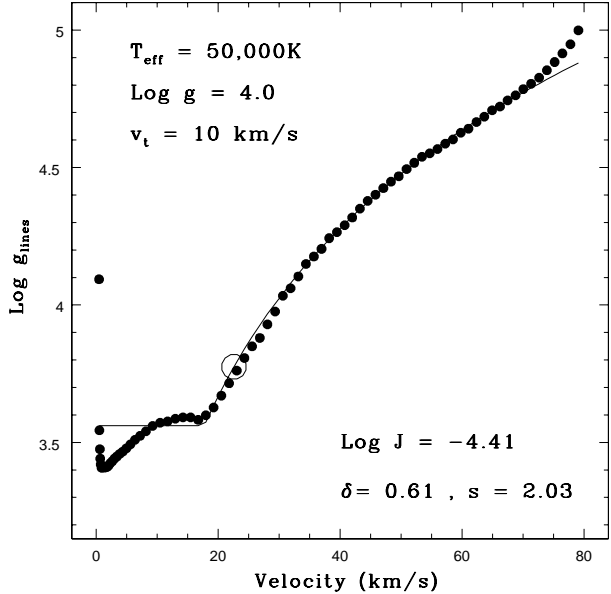}
\caption{Comparison of radiative accelerations due to lines. The
filled circles are values given by the MC estimator
$\tilde{g}_{\ell}$, and the solid line is the function $g_{\ell}(v)$ assumed in
deriving the reversing layer's stratification. The open circle indicates the
sonic point.}
\end{figure}

\subsection{Model D-40}  

Model D-40 of Pauldrach et al. has 
$T_{eff} = 40000$K, $log \: g = 3.75$, $R/R_{\sun} = 10$, and 
$\Phi = 0.24 \times 10^{-6} M_{\sun}/yr$, corresponding to 
$J(gm \: s^{-1})  = -5.60$ dex .
In this case, the source of 
b-values at the representative point and of the continuum-flux distribution is
the TLUSTY model G40000g375v10.12.

The best-fit moving reversing layer for $v_{t} = 10$ km s$^{-1}$ has 
$(log J, \delta, s) = (-5.65,0.58,2.55)$, with $\eta = 0.048$. Thus the mass
flux is lower
than that of the WM-{\em basic} model by a barely significant 0.05 dex.

\subsection{Model S-30}  

Model S-30 has 
$T_{eff} = 30000$K, $log \: g = 3.00$, $R/R_{\sun} = 27$, and 
$\Phi = 5.0 \times 10^{-6} M_{\sun}/yr$, corresponding to
$J(gm \: s^{-1})  = -5.15$ dex .
The matching TLUSTY model is G30000g300v10.12.

The best-fit moving reversing layer for $v_{t} = 10$ km s$^{-1}$ has 
$(log J, \delta, s) = (-5.45,0.79,1.75)$, with $\eta = 0.072$.
Thus the mass flux is lower
than that of the WM-{\em basic} model by an apparently significant 0.3 dex. 
However, the reliability of this model has been questioned
(Puls et al. 2005, p686).

\subsection{Consistency checks}  

The good agreement of the $J$'s with those of the reliable  
WM-{\em basic} models provides strong support for the simplifying assumptions
introduced in Sects. 2 \& 3.
A further consistency check is obtained from
the lines contributing to $\tilde{g}_{\ell}$. 
The treatment of ionization and excitation in Sect. 2.5 focuses on low-lying
energy levels in the expectation that the bulk of the radiative driving
derives from scattering by such levels. This is tested for the layer with
$v \simeq a$ in model D-50. For this layer, $\tilde{g}_{\ell} \simeq g_{*}$,
with the fractions 0.70 and 0.88 arising from levels with excitation
energies $< 5$ and $< 10$ eV, respectively.

\section{Variation of parameters}  

In this section, insight is sought into the behaviour of moving reversing
layers by varying parameters.

\subsection{Mass flux}

The excellent fit of $\tilde{g}_{\ell}$ to $g_{\ell}$ in Fig. 2 is
lost if parameters change.
In Fig.3,
the effects of the displacements $\Delta J = \pm 0.3$ dex are shown.
The changes in $\tilde{g}_{\ell}$ are in the expected sense. Thus an increase
in $J$ results in a $\tilde{g}_{\ell}$
that falls below the assumed $g_{\ell}(v)$ and is therefore incapable of 
accelerating the outflow.   

Fig. 3 supports the  
assumption of an X-type topology since there is no hint of the  
solution indeterminacy characteristic of a nodal-type topology (Poe et al
1990). The sensitivity to mass flux is consistent with there being a unique
eigenvalue $J$ for which there is a stationary outflow accelerating from sub-
to supersonic velocities.

\subsection{Microturbulence}

In a second sensitivity experiment, the effect of varying $v_{t}$
is investigated. Fig. 4 plots $\tilde{g}_{\ell}(v)$ for
$v_{t} = 6.7$ and $15$ km s$^{-1}$. In this case, the results are less readily
anticipated.

\begin{figure}
\vspace{8.2cm}
\includegraphics{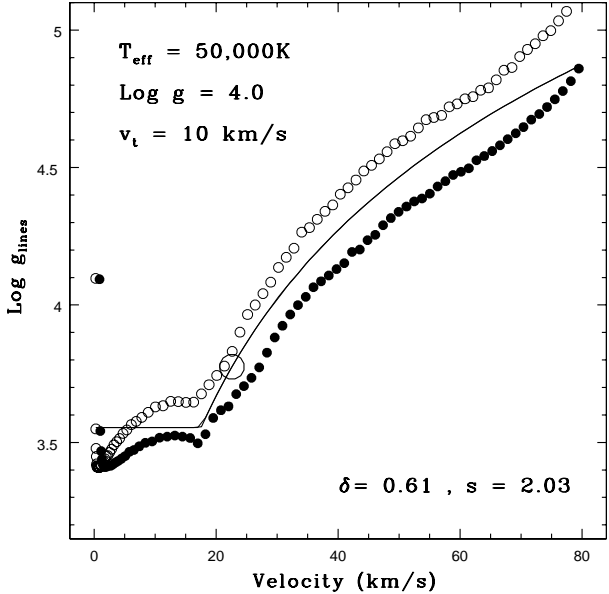}
\caption{Effect of varying mass flux $J$, with other
parameters as in Fig. 2. The filled and open circles are values of
$\tilde{g}_{\ell}$ when $J (gm \: s^{-1}) = -4.11$ and $-4.71$ dex,
respectively
The solid line is the function $g_{\ell}(v)$ assumed in
deriving the reversing layer's stratification, and the large open circle
indicates the sonic point.}
\end{figure}

For $v \la 15$ km s$^{-1}$, $\tilde{g}_{\ell}$ increases with $v_{t}$. This
is a consequence of the broadened lines in the quasi-static
layers intercepting an increased fraction of the continuum. A simple analysis
of this effect is given in Sect. IIb of LS70. With a slowly-varying
logarithmic factor neglected, this analysis shows that,
in a static atmosphere with a fixed column density of absorbing ions,
a strong resonance line has $g_{\ell} \propto \Delta \nu_{D}$, which is
$\propto v_{t}$ when microturbulence dominates the ions' thermal motions. 

Following a crossover at $v \simeq 20$ km s$^{-1}$, this dependency is reversed
and $\tilde{g}_{\ell}$ becomes an decreasing function of $v_{t}$. This is
again a consequence of the broadening of lines formed in the quasi-static
layers.
Continuum radiation that would have been available to accelerate
matter moving with $v \sim 30-40$ km s$^{-1}$ has already been partly absorbed 
in the turbulent flow at subsonic bulk velocities.
    
The final feature to be explained is the convergence of the
$\tilde{g}_{\ell}$ plots for $v \ga 60$ km s$^{-1}$. This is the Sobolev
regime. Matter has been accelerated from out of the shadow of lines formed in
the quasi-static flow and the narrow-line limit is approached in which
$g_{\ell}$ becomes independent of $\phi_{\nu}$ and therefore of $v_{t}$. 

With Fig. 4 understood, the next point of interest is the 
change in $(J, \delta, s)$ required to recover a good
fit between $\tilde{g}_{\ell}$ and $g_{\ell}$.
The above discussion indicates that as $v_{t}$ increases
so will $\delta$, as lines then provide more support for the quasi-static
layers,
but $J$ will decrease to compensate for the reduced flux available to
accelerate matter through the sonic point.
 
For each of the three WM-{\em basic} models considered in Sect. 4,
a sequence of best-fit models with varying $v_{t}$ is given in Table 1.
Each sequence has the 
continuum flux distribution and the b-values of the
TLUSTY model identified in Sects. 4.2-4.4. In consequence, there is an
inconsistency since the TLUSTY models assume $v_{t} = 10$ km s$^{-1}$. But the
dominant effects of $v_{t}$ on the parameter vector are surely
those just discussed. Note that the effect of 
$v_{t}$ on line-blocking and hence backwarming is approximately accounted for
by Eq. (7).
Thus, because packets that re-enter the photosphere are
implicitly thermalized and re-radiated, the continuum flux at $x_{0}$
exceeds $\sigma T_{eff}^{4}$ by an amount that depends on $v_{t}$.

From Table 1, we see that $\delta$ and $J$ vary in the sense expected.
In particular, $J$ decreases by $\simeq 0.3$dex as $v_{t}$ increases by
0.18dex,
corresponding to a power-law exponent of $-1.7$.
However, although this effect is readily understood and
undoubtedly real, the quality of the fits is markedly poorer for the D-50
models with $v_{t} = 15$ and $22.5$ km s$^{-1}$. This indicates the
need for a more complicated fitting function $g_{\ell}(v)$ to replace Eq. (4).

Turbulent broadening is treated here in the microturbulent limit in which the
length scale of random velocity fluctuations is $\ll$ the mean free
paths of line photons.
But the mechanism does not require this assumption: any
photospheric line-broadening that reduces $\rho g_{\ell}$ at
$v \simeq a$ will contribute to this throttling effect. In particular,
therefore, turbulence with
length scales up to $\sim$ the photospheric scale height might well be
important, and there is strong observational evidence
(Conti \& Ebbets 1977; Howarth et al. 1997) for turbulent
velocities well in excess of the standard $v_{t} \simeq 10$ km s$^{-1}$
required by line strengths.

\begin{table}

\caption{Models with varying microturbulence}

\label{table:1}

\centering

\begin{tabular}{c c c c c c}

\hline\hline

$Model$ &  $v_{t}$  &  $\delta$ &  $s$  &  $log J$  &  $\eta \times 100$ \\

\hline
\hline

D-50  &  6.7   &  0.48   &  1.96  &  -4.09  &  2.6  \\

      & 10.0   &  0.61   &  2.03  &  -4.41  &  4.8  \\

      & 15.0   &  0.69   &  1.59  &  -4.71  & 10.3  \\

      & 22.5   &  0.80   &  1.15  &  -5.00  & 10.4  \\

\cline{1-6}

D-40  &  6.7   &  0.52   &  2.30  &  -5.34  &  8.5  \\

      & 10.0   &  0.58   &  2.55  &  -5.65  &  4.8  \\

\cline{1-6}

S-30  &  6.7   &  0.74   &  2.20  &  -5.15  &  6.1  \\

      & 10.0   &  0.79   &  1.75  &  -5.45  &  7.2  \\

\hline
\hline 

\end{tabular}

\end{table}

\begin{figure}
\vspace{8.2cm}
\includegraphics{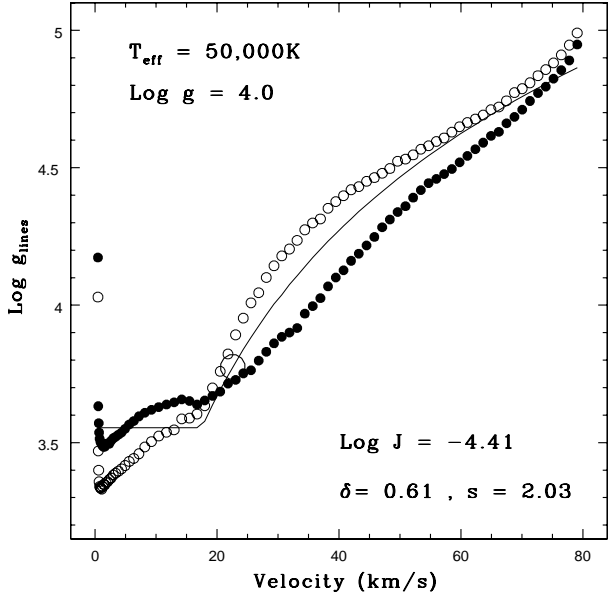}
\caption{Effect of varying microturbulent velocity $v_{t}$, with other
parameters as in Fig. 2. The filled and open circles are values of
$\tilde{g}_{\ell}$ when $v_{t} = 15$ and $6.7$ km s$^{-1}$, respectively
The solid line is the function $g_{\ell}(v)$ assumed in
deriving the reversing layer's stratification, and the large open circle
indicates
the sonic point.}
\end{figure}

\section{Discussion}

The above calculations reveal a fundamental coupling between line
formation in the quasi-static photospheric layers and the mass flux $J$ that
can be accelerated to supersonic velocities. This low-velocity, non-Sobolev
effect must be accurately treated when $J$ or $\Phi$ are computed as
eigenvalues of the equations of stationary line-driven winds.

\subsection{Previous work}  

This coupling was already recognized in LS70. In Fig.2 of that paper, the
precipitous drop of $J$ at $log T_{eff} \simeq 4.5$ was attributed to the
absorption of radiation at $\nu_{i} (1+a/c)$ by the damping wings of the
C$\:${\sc iv} doublet. Here, with the inclusion of numerous weak lines,
$J$ falls less precipitously when photospheric line broadening reduces
the contributions of strong lines to $\rho g_{\ell}$ at
$v \simeq a$. Moreover, rather than damping wings, turbulent broadening
of Doppler cores is a more plausible cause of a reduced radiation force in the
transonic flow.

Because the throttling effect of turbulent broadening
requires non-Sobolev line transfer, it is absent from
CAK theory. The ion thermal speed $v_{th}$ that appears in
CAK formulae is for dimensional reasons and scales out when
derived properties are expressed in physical units 
(Abbott 1982, Poe et al. 1990).
 
The hydrodynamic structure of WM-{\em basic} models are obtained with CAK
theory and thus a dependence of the eigenvalue
$\Phi$ on photospheric turbulence would seem to be excluded.
However, the CAK force multiplier
parameters are iteratively adjusted to fit $g_{\ell}$ given
by a detailed NLTE transfer code that assumes a velocity-dependent 
microturbulence
with $v_{t} = 10$ km s$^{-1}$ at the photosphere. Although this introduces
a dependence of $\Phi$ on $v_{t}$, the displacement of the CAK 
critical point into the supersonic flow suggests that
the throttling effect
will be less powerful than found here. But if the CAK formalism were eliminated
and the stratification derived directly from the gas dynamic equations
with critical point at the sonic point, an accurate treatment of the 
dependence of $\Phi$ on $v_{t}$ would be possible with this code.

Although not identified with turbulence,
the throttling effect of an enhanced $v_{th}/a$ is evident in the pure
absorption calculations reported by Poe et al. (1990). Moreover, they
summarize the implications of this parameter for time-dependent simulations
and the uniqueness of stationary solutions.

Hillier et al. (2003) identify the treatment of microturbulence in 
calculating $\rho g_{\ell}$ near the sonic point as a major unsolved issue
and recognize that the Poe et al. work can be re-interpreted to anticipate the
effects of microturbulence.

\subsection{Calculating $\Phi$}  

Two theoretical techniques for computing $\Phi$ have emerged, one
fundamental, the other semi-empirical. The fundamental approach determines
$\Phi$ or $J$ as an eigenvalue of the governing differential equations
by demanding regularity at the critical point. The
semi-empirical approach, on the other hand,
determines $\Phi$ algebraically by demanding that the flux of mechanical
energy in the terminal flow should be accounted for by the loss of radiative
luminosity due to line- and electron scatterings in the outflowing matter. 

A consequence of this investigation is that the fundamental approach is
harder than hitherto realized. An essential phenomenon, photospheric
turbulence,
is not predictable as yet from first principles and only 
measurable with moderate accuracy, even in the favourable but unlikely
circumstances that such
motions are restricted to the microturbulent
limit and do not vary with height. Accordingly, by default, the semi-empirical
technique is currently preferable as a source of $\Phi$'s for input into other 
astrophysical investigations. The details of transonic
flow are then inconsequential because
the kinetic energy density at the sonic point is negligible compared to
that of the terminal flow ($v_{\infty}^{2}/2$) which, moreover, is an
observable.
  
As discussed in Sect. 1, the semi-empirical $\Phi$'s obtained by Vink et al.
(2000) are too high according to Bouret et al. (2005). This may well be
because the AL85 assumptions of laminar flow with a
monotonic velocity law are retained. But these assumptions
are not fundamental to the MC technique. The Vink et al. (2000) investigation
could usefully be repeated with the stratification updated in accordance with
the already cited papers on clumpiness as well as the recent papers
of Puls et al. (2006) and Oskinova et al. (2006). Moreover, the treatment of
line formation could be improved to treat branching (Sim 2004).

A survey of $\Phi$'s for hot stars could
also be carried out with the diagnostic code CMFGEN, which already 
incorporates clumpiness and random non-monoticity. For radii
where
the bulk velocity $v(r)$ is highly supersonic, we have 
to a good approximation  
\begin{equation}
   \frac{1}{2} v^{2}= \int_{R}^{r} (g_{\ell} - g_{*}) dr
\end{equation}
By demanding that this equation is satisfied at $r = \infty$, $\Phi$ could be
determined algebraically in a manner analogous to AL85 and Vink et al. (2000).
Further wind parameters, such as $\beta$ in the velocity law, could be
determined by minimizing
violations of this equation over its range of validity. 

\section{Conclusion}

The aim of this paper has been to examine claims that the theory of
line-driven winds overpredicts mass loss rates. To this end, the theory of
moving reversing layers has been updated and used to investigate the
sensitivity of transonic flow to input parameters. The reported calculations
reveal that photospheric turbulence, by reducing the driving force at the
sonic point, plays a decisive role in regulating the mass flux of stationary
flow. Codes that
compute $\Phi$ as an eigenvalue will therefore overpredict if turbulent
line-broadening is underestimated. This may occur as a result of
observational underestimates of $v_{t}$ or because 
the microturbulent limit is violated. A further possible cause of
overprediction is reliance on the CAK critical point in determining the
eigenvalue. Because this point is at supersonic velocity, the throttling
effect of photospheric turbulence is likely to be reduced, a topic meriting
further research.

The calculations presented in Table 1 suggest the conjecture that the mass
loss rates
of hot stars are less than predicted because of the
throttling effect of photospheric turbulence.
With the critical point at the sonic point as required by non-Sobolev 
line transfer,
this effect reduces $J$ for stationary flow by a factor $\ga 3$ for near
sonic turbulence.

However, additional work is required to establish this conjecture. The
MC technique presented herein is barely adequate for computing $J$'s when 
$v_{t} \ga 15$ km s$^{-1}$. Repetition with an improved parameterization of
$g_{\ell}$ or solution with an alternative procedure is therefore desirable.
But, more fundamentally, the conjecture is motivated by the behaviour of 
stationary solutions, and these are unstable (Sect.1). Ultimately, therefore,
as the referee remarks, time-dependent 
calculations are needed to see if the time-averaged $J$ is similarly
constrained by photosperic turbulence.

With regard to analyses of individual stars,
the eigenvalue $\Phi = 4 \pi R^{2} J$ obtained from a moving reversing layer
computed with the
spectroscopic $v_{t}$ should be compared with the empirical $\Phi$
obtained from the supersonic wind with clumpiness and non-monoticity included. 
Such analyses could usefully exploit the wide range of $v_{t}$'s reported for
O stars (Bouret et al. 2003; Heap et al. 2006).

\acknowledgement

I am grateful to the referee, J.Puls, for detailed comments on the original
version.


\begin{thebibliography}{}

\bibitem[]{}
 
 Abbott, D.C. 1982,ApJ, 259, 282

\bibitem[]{}

 Abbott, D.C., Bieging, J.H., \& Churchwell, E. 1981, ApJ, 250, 645

\bibitem[]{}

 Abbott, D.C., \& Lucy, L.B. 1985, ApJ, 288, 679 (AL85)

\bibitem[]{}

 Bouret, J.-C., Lanz, T., Hillier, D. J., Heap, S. R., Hubeny, I.,
            Lennon, D. J., Smith, L. J., \& Evans, C. J. 2003, ApJ, 595, 1182

\bibitem[]{}

 Bouret, J.-C., Lanz, T., \& Hillier, D. J. 2005, A\&A, 438, 301


\bibitem[]{}

 Castor, J.I., Abbott, D.C., \& Klein, R.I. 1975, ApJ, 195, 157 (CAK)

\bibitem[]{}

 Conti, P. S., \& Ebbets, D.   1977, ApJ, 213, 438
     
\bibitem[]{}

 Fullerton, A.W., Massa, D.L., \& Prinja, R.K. 2006, ApJ, 637, 1025

\bibitem[]{}

 Grevess, N., \& Sauval, A.J. 1998, Sp. Sci. Rev.,85,161

\bibitem[]{}

 Heap, S. R., Lanz, T., \& Hubeny, I. 2006, ApJ, 638, 409

\bibitem[]{}

 Hillier, D.J., \& Miller, D.L. 1998, ApJ, 496, 407

\bibitem[]{}

 Hillier, D.J., \& Miller, D.L. 1999, ApJ, 519, 345

\bibitem[]{}

 Hillier, D.J., Lanz, T., Heap, S. R., Hubeny, I., Smith, L. J., Evans, C. J.,
           Lennon, D. J., \& Bouret, J. C. 2003, ApJ,588,1039

\bibitem[]{}

 Howarth, I. D., Siebert, K. W., Hussain, G. A. J., \& Prinja, R. K.
	   1997, MNRAS, 284, 265


\bibitem[]{}

 Rybicki, G. B., \& Hummer, D. G. 1969, MNRAS, 144, 313

\bibitem[]{}

 Kurucz, R.L., \& Bell, B. 1995, Kurucz CD-ROM No.23

\bibitem[]{}

 Lanz, T. \& Hubeny, I. 2003, ApJS, 146, 417

\bibitem[]{}

 Lucy, L.B. \& Solomon, P.M. 1970, ApJ, 159, 879 (LS70)

\bibitem[]{}

 Lucy, L.B. 1975, Mem.Soc.R.Sci.Li\`{e}ge,8,359

\bibitem[]{}

 Lucy, L.B. 2005, A\&A, 429, 19 

\bibitem[]{}

 Oskinova, L. M., Feldmeier, A., \& Hamann, W.-R. 2006, MNRAS, 372, 313

\bibitem[]{}

 Owocki, S.P., Castor, J.I., \& Rybicki, G.B. 1988, ApJ, 335, 914

\bibitem[]{}

 Owocki, S. P. \& Puls, J. 1999, ApJ, 510, 355

\bibitem[]{}

 Pauldrach,A.W.A., Hoffmann,T.L., \& Lennon,M. 2001, A\&A, 375, 161 

\bibitem[]{}

 Poe,C.H., Owocki,S.P., \& Castor,J.I. 1990, ApJ, 358, 199

\bibitem[]{}

 Press W.H., Teukolsky S.A., Vetterling W.T., Flannery B.P. 1992, Numerical
          Recipes. Cambridge Univ. Press, Cambridge  

\bibitem[]{}

 Puls,J., Urbaneja,M.A., Venero,R., Repolust,T., Springmann,U., Jokuthy,A.,
             \& Mokiem,M.R. 2005, A\&A, 435, 669	


\bibitem[]{}

 Puls,J., Markova,N., Scuderi,S., Stanghellini,C., Taranova,O.G., 
             Burnley,A.W., \& Howarth,I.D. 2006, A\&A, 454, 625	

\bibitem[]{}

 Schmutz,W. 1997, A\&A, 321, 268

\bibitem[]{}

 Sim, S. 2004, MNRAS, 349, 899


\bibitem[]{}

 Vink, J.S., de Koter, A., \& Lamers, H.J.G.L.M. 2000, A\&A, 362, 295


\end{thebibliography}
\end{document}